\documentstyle[11pt,epsfig]{article}

\begin{document}

\title{Decay Constants of Heavy Vector Mesons \\
in Relativistic Bethe-Salpeter Method} \vspace{10mm}

\author{
Guo-Li~ Wang \footnote{gl\_wang@hit.edu.cn} \\{\it \small
Department of Physics, Harbin Institute of Technology, Harbin,
150006, China.}}

\date{}

\maketitle

\baselineskip=24pt
\begin{quotation}
\vspace*{1.5cm}
\begin{center}
  \begin{bf}
  ABSTRACT
  \end{bf}
\end{center}

\vspace*{0.5cm} \noindent In a previous letter, we computed the
decay constants of heavy pseudoscalar mesons in the framework of
relativistic (instantaneous) Bethe-Salpeter method (full $0^-$
Salpeter equation), in this letter, we solve the full $1^-$
Salpeter equation and compute the leptonic decay constants of
heavy-heavy and heavy-light vector mesons. The theoretical
estimate of mass spectra of these heavy-heavy and heavy-light
vector mesons are also presented. Our results for the decay
constants and mass spectra include the complete relativistic
contributions. We find $F_{D^*_s} \approx~ 375~\pm ~24~$, $F_{D^*}
\approx~ 340~\pm ~23~ (D^{*0},D^{*\pm})$, $F_{B^*_s} \approx~
272~\pm ~20~$, $F_{B^*} \approx~ 238~\pm ~18~ (B^{*0},B^{*\pm})$,
$F_{B^*_c} \approx~ 418~\pm ~24~$, $F_{J/\Psi} \approx~ 459~ \pm
~28~$, $F_{\Psi(2S)} \approx~ 364~ \pm ~24~$, $F_{\Upsilon}
\approx~ 498~\pm ~20~$ and $F_{\Upsilon(2S)} \approx~ 366~\pm
~27~$ MeV.

\end{quotation}

\newpage
 \setcounter{page}{1}
\section{Introduction}

     The decay constants of mesons are very important quantities
\cite{Adler,Pagels,Reinders,Allton} and the study of them has
become an interesting topic in recent years
\cite{fp,bes,cleo2,beatrice,opal,aleph,Roberts,narison,penin,yamada,ryan,bernard,juttner,wangzg}.
These constants play important roles in many aspects, such as in
the determination of Cabibbo-Kobayashi-Maskawa matrix elements, in
the leptonic or nonleptonic weak decays of mesons, in the neutral
$D-\bar{D}$ or $B-\bar{B}$ mixing process, etc.

In a previous letter \cite{previous}, the decay constants of
heavy-heavy and heavy-light pseudoscalar mesons are calculated in
the framework of relativistic instantaneous Bethe-Salpeter method
\cite{BS} (also called Salpeter method \cite{salp}), good
agreement of our predictions with recent lattice, QCD sum rule,
other relativistic model calculations as well as available
experimental data is found.

In this letter, we extend our previous analysis to include vector
mesons, present the relativistic calculation of heavy-heavy and
heavy-light vector decay constants in the framework of full
Salpeter equation. The instantaneous Bethe-Salpeter equation,
which is also called full Salpeter equation, is a relativistic
equation describing a bound state. Since this method has a very
solid basis in quantum field theory, it is very good in describing
a bound state which is a relativistic system. In a earlier paper
\cite{cskimwang}, we solved the full $0^{-}$ Salpeter equation of
pseudoscalar mesons; in another paper \cite{changwang}, we solved
the full $1^{--}$ Salpeter equation of equal-mass vector mesons.
The predictions of these relativistic methods agree very well with
other theoretical calculation and recent experimental data. In
this letter, we extend the analysis of equal-mass vector system to
non-equal mass system by solving the full $1^{-}$ Salpeter
equation of vector mesons, and use this method to predict the
values of decay constants for heavy-heavy and heavy-light vector
mesons. There are some input parameters in our method, we need to
fix them when we solve the full $1^-$ Salpeter equation. In our
calculation, we fix the input parameters by fitting the
experimental data of mass spectra, so we also present the mass
spectra for heavy vector mesons.

This letter is organized as following, in section 2, we introduce
the relativistic Bethe-Salpeter equation and Salpeter equation. In
section 3, we give the formula of relativistic wave function and
decay constants of vector meson. We solve the full Salpeter
equation, obtain the mass spectra and wave function of vector
mesons. Finally, we use these relativistic wave function to
calculate the decay constants of heavy vector mesons and show the
numerical results and conclusion in section 4.

\section{Instantaneous Bethe-Salpeter Equation}

In this section, we briefly review the instantaneous
Bethe-Salpeter equation and introduce our notations, interested
reader can find the details in Ref. \cite{cskimwang}.

The Bethe-Salpeter (BS) equation is read as \cite{BS}:
\begin{equation}
(\not\!{p_{1}}-m_{1})\chi(q)(\not\!{p_{2}}+m_{2})=
i\int\frac{d^{4}k}{(2\pi)^{4}}V(P,k,q)\chi(k)\;, \label{eq1}
\end{equation}
where $\chi(q)$ is the BS wave function with the total momentum
$P$ and relative momentum $q$ of the bound state, and $V(P,k,q)$
is the kernel between the quarks in the bound state. $p_{1},
p_{2}$ are the momenta of the constituent quark 1 (heavy or light)
and anti-quark 2 (heavy or light), respectively. The total
momentum $P$ and the relative momentum $q$ are defined as:
$$p_{1}={\alpha}_{1}P+q, \;\; {\alpha}_{1}=\frac{m_{1}}{m_{1}+m_{2}}~,$$
$$p_{2}={\alpha}_{2}P-q, \;\; {\alpha}_{2}=\frac{m_{2}}{m_{1}+m_{2}}~.$$

The BS wave function $\chi(q)$ satisfy the following normalization
condition:
\begin{equation}
\int\frac{d^{4}k d^{4}q} {(2\pi)^{4}}Tr\left[\overline\chi(k)
\frac{\partial}{\partial{P_{0}}}\left[S_{1}^{-1}(p_{1})
S_{2}^{-1}(p_{2})\delta^{4}(k-q)+
V(P,k,q)\right]\chi(q)\right]=2iP_{0}\;, \label{eq2}
\end{equation}
where $S_1(p_{1})$ and $S_2(p_{2})$ are the propagators of the two
constituents. In many applications, the kernel of BS equation is
``instantaneous'', $i.e.$, in the center mass frame of the
concerned bound state ($\stackrel{\rightarrow}{P}=0$), the kernel
$V(P,k,q)$ takes the simple form:
$$V(P,k,q) \Rightarrow V(k,q)=V(|\stackrel{\rightarrow}{k}|,
|\stackrel{\rightarrow}{q}|, \cos\theta)\;,$$ where $\theta$ is
the angle between the vectors $\stackrel{\rightarrow}{k}$ and
$\stackrel{\rightarrow}{q}$. Then the BS equation reduced to the
Salpeter equation.

It is convenient to divide the relative momentum $q$ into two
parts, $q_{\parallel}$ and $q_{\perp}$, a parallel part and an
orthogonal one to the total momentum of the bound state,
respectively.
$$q^{\mu}=q^{\mu}_{\parallel}+q^{\mu}_{\perp}\;,$$
$$q^{\mu}_{\parallel}\equiv (P\cdot q/M^{2})P^{\mu}\;,\;\;\;
q^{\mu}_{\perp}\equiv q^{\mu}-q^{\mu}_{\parallel}\;.$$
Correspondingly, we have two Lorentz invariant variables:
\begin{center}
$q_{p}=\frac{(P\cdot q)}{M}\;, \;\;\;\;\;
q_{_T}=\sqrt{q^{2}_{p}-q^{2}}=\sqrt{-q^{2}_{\perp}}\;.$
\end{center}
In the center of mass frame $\stackrel{\rightarrow}{P}=0$, they
turn out to the usual component $q_{0}$ and
$|\stackrel{\rightarrow}{q}|$, respectively. Now the volume
element of the relative momentum $k$ can be written in an
invariant form:
\begin{equation}
d^{4}k=dk_{p}k^{2}_{T}dk_{_T}ds d\phi\;, \label{eq3}
\end{equation}
where $\phi$ is the azimuthal angle, $s=(k_{p}q_{p}-k\cdot
q)/(k_{_T}q_{_T})$. The instantaneous interaction kernel can be
rewritten as:
\begin{equation}
V(|\stackrel{\rightarrow}{k}-\stackrel{\rightarrow}{q}|)=
V(k_{\perp},q_{\perp},s)\;. \label{eq4}
\end{equation}

Let us introduce the notations $\varphi_{p}(q^{\mu}_{\perp})$ and
$\eta(q^{\mu}_{\perp})$ for three dimensional wave function as
follows:
$$
\varphi_{p}(q^{\mu}_{\perp})\equiv i\int
\frac{dq_{p}}{2\pi}\chi(q^{\mu}_{\parallel},q^{\mu}_{\perp})\;,
$$
\begin{equation}
\eta(q^{\mu}_{\perp})\equiv\int\frac{k^{2}_{_T}dk_{_T}ds}{(2\pi)^{2}}
V(k_{\perp},q_{\perp},s)\varphi_{p}(k^{\mu}_{\perp})\;.
\label{eq5}
\end{equation}
Then the BS equation can be rewritten as:
\begin{equation}
\chi(q_{\parallel},q_{\perp})=S_{1}(p_{1})\eta(q_{\perp})S_{2}(p_{2})\;.
\label{eq6}
\end{equation}
The propagators of the two constituents can be decomposed as:
\begin{equation}
S_{i}(p_{i})=\frac{\Lambda^{+}_{ip}(q_{\perp})}{J(i)q_{p}
+\alpha_{i}M-\omega_{i}+i\epsilon}+
\frac{\Lambda^{-}_{ip}(q_{\perp})}{J(i)q_{p}+\alpha_{i}M+\omega_{i}-i\epsilon}\;,
\label{eq7}
\end{equation}
with
\begin{equation}
\omega_{i}=\sqrt{m_{i}^{2}+q^{2}_{_T}}\;,\;\;\;
\Lambda^{\pm}_{ip}(q_{\perp})= \frac{1}{2\omega_{ip}}\left[
\frac{\not\!{P}}{M}\omega_{i}\pm
J(i)(m_{i}+{\not\!q}_{\perp})\right]\;, \label{eq8}
\end{equation}
where $i=1, 2$ for quark and anti-quark, respectively,
 and
$J(i)=(-1)^{i+1}$. Here $\Lambda^{\pm}_{ip}(q_{\perp})$ satisfy
the relations:
\begin{equation}
\Lambda^{+}_{ip}(q_{\perp})+\Lambda^{-}_{ip}(q_{\perp})=\frac{\not\!{P}}{M}~,\;\;
\Lambda^{\pm}_{ip}(q_{\perp})\frac{\not\!{P}}{M}
\Lambda^{\pm}_{ip}(q_{\perp})=\Lambda^{\pm}_{ip}(q_{\perp})~,\;\;
\Lambda^{\pm}_{ip}(q_{\perp})\frac{\not\!{P}}{M}
\Lambda^{\mp}_{ip}(q_{\perp})=0~. \label{eq9}
\end{equation}
Due to these equations, $ \Lambda^{\pm}$ may be considered as
$P-$projection operators, since in the rest frame
$\overrightarrow{P}=0$ they turn to the energy projection
operator.

Introducing the notations $\varphi^{\pm\pm}_{p}(q_{\perp})$ as:
\begin{equation}
\varphi^{\pm\pm}_{p}(q_{\perp})\equiv
\Lambda^{\pm}_{1p}(q_{\perp})
\frac{\not\!{P}}{M}\varphi_{p}(q_{\perp}) \frac{\not\!{P}}{M}
\Lambda^{{\pm}}_{2p}(q_{\perp})\;, \label{eq10}
\end{equation}
and taking into account with
$\frac{\not\!{P}}{M}\frac{\not\!{P}}{M}=1$, we have
$$
\varphi_{p}(q_{\perp})=\varphi^{++}_{p}(q_{\perp})+
\varphi^{+-}_{p}(q_{\perp})+\varphi^{-+}_{p}(q_{\perp})
+\varphi^{--}_{p}(q_{\perp})
$$
With contour integration over $q_{p}$ on both sides of
Eq.(\ref{eq6}), we obtain:
$$
\varphi_{p}(q_{\perp})=\frac{
\Lambda^{+}_{1p}(q_{\perp})\eta_{p}(q_{\perp})\Lambda^{+}_{2p}(q_{\perp})}
{(M-\omega_{1}-\omega_{2})}- \frac{
\Lambda^{-}_{1p}(q_{\perp})\eta_{p}(q_{\perp})\Lambda^{-}_{2p}(q_{\perp})}
{(M+\omega_{1}+\omega_{2})}\;,
$$
and we may decompose it further into four equations as follows:
$$
(M-\omega_{1}-\omega_{2})\varphi^{++}_{p}(q_{\perp})=
\Lambda^{+}_{1p}(q_{\perp})\eta_{p}(q_{\perp})\Lambda^{+}_{2p}(q_{\perp})\;,
$$
$$(M+\omega_{1}+\omega_{2})\varphi^{--}_{p}(q_{\perp})=-
\Lambda^{-}_{1p}(q_{\perp})\eta_{p}(q_{\perp})\Lambda^{-}_{2p}(q_{\perp})\;,$$
\begin{equation}
\varphi^{+-}_{p}(q_{\perp})=\varphi^{-+}_{p}(q_{\perp})=0\;.
\label{eq11}
\end{equation}

The complete normalization condition (keep all of the four
components appearing in Eq.(\ref{eq11})) for BS equation turns out
to be:
\begin{equation}
\int\frac{q_{_T}^2dq_{_T}}{2{\pi}^2}Tr\left[\overline\varphi^{++}
\frac{{/}\!\!\!
{P}}{M}\varphi^{++}\frac{{/}\!\!\!{P}}{M}-\overline\varphi^{--}
\frac{{/}\!\!\! {P}}{M}\varphi^{--}\frac{{/}\!\!\!
{P}}{M}\right]=2P_{0}\;. \label{eq12}
\end{equation}

\section{Relativistic Wave Function and Decay Constant of Vector Meson}

The general form for the relativistic wave function of vector
state $J^P=1^{-}$ can be written as 16 terms constructed by $P$,
$q$, $\epsilon$ and gamma matrix. Because of the approximation of
instantaneous, the 8 terms with $P\cdot q_{\perp}$ become zero, so
the general form for the relativistic Salpeter wave function of
vector state $J^P=1^{-}$ can be written as \cite{changwang}:
$$\varphi_{1^{-}}^{\lambda}(q_{\perp})=
q_{\perp}\cdot{\epsilon}^{\lambda}_{\perp}
\left[f_1(q_{\perp})+\frac{\not\!P}{M}f_2(q_{\perp})+
\frac{{\not\!q}_{\perp}}{M}f_3(q_{\perp})+\frac{{\not\!P}
{\not\!q}_{\perp}}{M^2} f_4(q_{\perp})\right]+
M{\not\!\epsilon}^{\lambda}_{\perp}f_5(q_{\perp})$$
\begin{equation}+
{\not\!\epsilon}^{\lambda}_{\perp}{\not\!P}f_6(q_{\perp})+
({\not\!q}_{\perp}{\not\!\epsilon}^{\lambda}_{\perp}-
q_{\perp}\cdot{\epsilon}^{\lambda}_{\perp})
f_7(q_{\perp})+\frac{1}{M}({\not\!P}{\not\!\epsilon}^{\lambda}_{\perp}
{\not\!q}_{\perp}-{\not\!P}q_{\perp}\cdot{\epsilon}^{\lambda}_{\perp})
f_8(q_{\perp}),\label{eq13}
\end{equation}
where the ${\epsilon}^{\lambda}_{\perp}$ is the polarization
vector of the vector meson. The equations
\begin{equation}
\varphi^{+-}_{1^{-}}(q_{\perp})=\varphi^{-+}_{1^{-}}(q_{\perp})=0\;
\end{equation}
give the constraints on the components of the wave function, so we
have the relations
$$f_1(q_{\perp})=\frac{\left[q_{\perp}^2 f_3(q_{\perp})+M^2f_5(q_{\perp})
\right] (m_1m_2-\omega_1\omega_2+q_{\perp}^2)}
{M(m_1+m_2)q_{\perp}^2},~~~f_7(q_{\perp})=\frac{f_5(q_{\perp})M(-m_1m_2+\omega_1\omega_2+q_{\perp}^2)}
{(m_1-m_2)q_{\perp}^2},$$
$$f_2(q_{\perp})=\frac{\left[-q_{\perp}^2 f_4(q_{\perp})+M^2f_6(q_{\perp})\right]
(m_1\omega_2-m_2\omega_1)}
{M(\omega_1+\omega_2)q_{\perp}^2},~~~f_8(q_{\perp})=\frac{f_6(q_{\perp})M(m_1\omega_2-m_2\omega_1)}
{(\omega_1-\omega_2)q_{\perp}^2}.$$ Then there are only four
independent wave functions $f_3(q_{\perp})$, $f_4(q_{\perp})$,
$f_5(q_{\perp})$ and $f_6(q_{\perp})$ been left in the
Eq.(\ref{eq13}). Following the Ref.\cite{cskimwang}, put
Eq.(\ref{eq13}) into Eq.(\ref{eq11}) and take trace, we obtain
four coupled integral equations, by solving them we obtain the
numerical results of mass spectra and wave functions for the
corresponding bound states.

In our calculation, we choose the center-of-mass system of the
heavy meson, so $q_{\parallel}$ and $q_{\perp}$ turn out to be the
usual components $(q_0, {\vec 0})$ and $(0,{\vec q})$,
$\omega_{1}=(m_{1}^{2}+{\vec q}^{2})^{1/2}$ and
$\omega_{2}=(m_{2}^{2}+{\vec q}^{2})^{1/2}$. Wave functions
$f_3({\vec q})$, $f_4({\vec q})$, $f_5({\vec q})$ and $f_6({\vec
q})$ will fulfill the normalization condition:
 \begin{equation}
\int \frac{d{\vec q}}{(2\pi)^3}\frac{16\omega_1\omega_2}{3}\left\{
3f_5f_6\frac{M^2}{m_1\omega_2+m_2\omega_1}+\frac{\omega_1\omega_2-m_1m_2+{\vec
q}^2}{(m_1+m_2)(\omega_1+\omega_2)}\left[
f_4f_5-f_3\left(f_4\frac{{\vec q}^2}{M^2}+f_6\right)\right]
\right\}=2M.
 \end{equation}

 In our model, Cornell
potential, a linear scalar interaction plus a vector interaction
is chosen as the instantaneous interaction kernel $V$
\cite{cskimwang}:
$$V(\stackrel{\rightarrow}{q})=V_s(\stackrel{\rightarrow}{q})
+\gamma_{_0}\otimes\gamma^0 V_v(\stackrel{\rightarrow}{q})~,$$
\begin{equation}
V_s(\stackrel{\rightarrow}{q})=-(\frac{\lambda}{\alpha}+V_0)
\delta^3(\stackrel{\rightarrow}{q})+\frac{\lambda}{\pi^2}
\frac{1}{{(\stackrel{\rightarrow}{q}}^2+{\alpha}^2)^2}~,
~~V_v(\stackrel{\rightarrow}{q})=-\frac{2}{3{\pi}^2}\frac{\alpha_s(
\stackrel{\rightarrow}{q})}{{(\stackrel{\rightarrow}{q}}^2+{\alpha}^2)}~,\label{eq16}
\end{equation}
where the coupling constant $\alpha_s(\stackrel{\rightarrow}{q})$
is running:
$$\alpha_s(\stackrel{\rightarrow}{q})=\frac{12\pi}{27}\frac{1}
{\log
(a+\frac{{\stackrel{\rightarrow}{q}}^2}{\Lambda^{2}_{QCD}})}~,$$
and the constants $\lambda$, $\alpha$, $a$, $V_0$ and
$\Lambda_{QCD}$ are the parameters that characterize the
potential.

The decay constant $F_V$ of vector meson is defined as
\begin{eqnarray}
\langle0|\bar{q_1}\gamma_\mu q_{2} |V,\epsilon\rangle &\equiv&
F_{V}M{\epsilon}^{\lambda}_\mu,
\end{eqnarray}
which can be written in the language of the Salpeter wave functions
as:
\begin{eqnarray}
\langle0|\bar{q_1}\gamma_\mu q_{2} |V,\epsilon\rangle &=&
\sqrt{N_c}\int Tr \left[\gamma_\mu \varphi_{1^-}({\vec q})
\frac{d{\vec q}}{(2\pi)^3} \right]=
4M\sqrt{N_c}{\epsilon}^{\lambda}_\mu \int \frac{d {\vec
q}}{(2\pi)^3} f_{5}({\vec q}) \label{fp}.
\end{eqnarray}
Therefore, we have
\begin{eqnarray}
F_{V} = 4\sqrt{N_c} \int \frac{d{\vec q}}{(2\pi)^3} f_{5}({\vec
q}),\label{eq19}\end{eqnarray}

\section{Numerical Results and Conclusion}
In our method, there are some parameters that have to be fixed
when performing the calculations. In Ref.~\cite{cskimwang}, we
fixed the values of the input parameters by fitting the mass
spectra for heavy pseudoscalar mesons of $0^-$ states, and we hope
to choose the same parameters for pseudoscalar and vector mesons.
But we find, when we solve the full Salpeter equation
Eq.(\ref{eq11}) of heavy vector mesons with same parameter set as
in Ref.~\cite{cskimwang} for pseudoscalar mesons, our predictions
of mass spectra of vector mesons do not agree well with
experimental data, and can not explain the vector-pseudoscalar
mass splitting. We argue that we choose a very simple interaction
kernel Eq.(\ref{eq16}), while the forms of pseudoscalar and vector
wave functions are very different (see Eq.(\ref{eq13}) in this
letter and Eq.(20) in Ref.\cite{cskimwang}). The latter decrease
the connection between the pseudoscalar and vector states, so we
like to choose different parameters to fit experimental data and
to present the mass splitting between the vector and pseudoscalar
mesons. We find the following parameters can fit data very well,
$$
a=e=2.7183~, ~~\alpha=0.06~ {\rm GeV}, ~~V_0=-0.49~ {\rm GeV},
~~\lambda=0.21~ {\rm GeV}^2, ~~\Lambda_{QCD}=0.27~ {\rm GeV}~~
{\rm and}
$$
\begin{equation} m_b=5.158~ {\rm GeV},~~ m_c=1.7551~ {\rm GeV},~~ m_s=0.535~ {\rm GeV},
m_d=0.377~ {\rm GeV},~~ m_u=0.371~ {\rm GeV}, \label
{para}\end{equation} and we show our predictions of mass spectra
for heavy vector mesons as well as the experimental data in Table
1 and Table 2.

In table 1, we show the results for ground state (1S) and first
radial excitation state (2S), one can see that our mass
predictions for ground states of heavy-light mesons can fit the
experimental data \cite{PDG} very well. In table 2, we show the
mass spectra of the first eight states for vector $c \bar c$
system. As can be seen from the table, our mass results below $2D$
state agree well with experimental data, while the masses for $2D$
and $4S$ states are about $50\sim60$ MeV lower than the
experimental data.
\begin{table}[]\begin{center}
\caption{Mass spectra in unit of $MeV$ for heavy vector meson.
`Ex' means the data from experiments \cite{PDG}. `Th' means the
predictions from our theoretical estimate.} \vspace{0.5cm}
\begin{tabular}
{|c|c|c|c|c|c|c|c|c|c|}\hline &$B^*_c$&$B^*_s$&$B^*_d$&$B^*_u$
&$D^*_s$&$D^*_d$&$D^*_u$\\\hline Ex(1S)
&&5416.6&5325.0&5325.0&2112.1 &2010.0&2006.7\\\hline Th(1S)
&6336.9&5416.6&5326.2&5322.9
&2112.0&2010.2&2006.5\\
\hline Th(2S) &6918.5&5957.6&5842.3&5837.7
&2673.0&2545.9&2540.8\\\hline
\end{tabular}
\end{center}
\end{table}
\begin{table}[]\begin{center}
\caption{Mass spectra in unit of $MeV$ for $c\bar c$ vector
system.} \vspace{0.5cm}
\begin{tabular}
{|c|c|c|c|c|c|c|c|c|}\hline &1S&2S&1D&3S&2D&4S&3D&5S\\\hline Ex$(c
\bar c )$&3096.916&3686.093&3770.0& 4040&4159&4415&&\\\hline Th$(c
\bar c )$&3096.8&3690.9&3759.8&4065.2&4108.2&4344.2&4371.6&4567.2
\\\hline
\end{tabular}
\end{center}
\end{table}

We also calculate the mass spectra for vector $b \bar b$ system,
we find our prediction with upper parameter set Eq.(\ref{para})
can not fit experimental data. The reason is due to that in $b
\bar b$ system, there are double heavy $b$ quarks, and the flavor
$N_f=4$, so we have to choose a new set of parameters as well as
smaller value of coupling constant \cite{twophoton}. We change the
previous scale parameters to $\Lambda_{QCD}=0.21$ GeV,
$m_b=5.1242$ GeV, and other parameters are not changed. With this
set of parameters, the coupling constant at the scale of bottom
quark mass is $\alpha_s(m_b)=0.23$, and obtained the mass spectra
of $b \bar b$ systems. We show the numerical results and
experimental data in Table 3. One can see that our predictions can
fit the experimental data very well, even for higher states.
\begin{table}[]\begin{center}
\caption{Mass spectra in unit of $MeV$ for $b \bar b$ vector
system.} \vspace{0.5cm}
\begin{tabular}
{|c|c|c|c|c|c|c|c|c|}\hline &1S&2S&1D&3S&2D&4S&3D&5S\\\hline Ex$(b
\bar b )$ &9460.30&10023.26&&10355.2&&10580.0&&10865\\\hline Th$(b
\bar b )$&9460.3&10029&10130&10379&10438&10648&10690&10868
\\\hline
\end{tabular}
\end{center}
\end{table}

By fitting the mass spectra of heavy mesons, we fixed the
parameters and obtained the relativistic wave functions for heavy
mesons. Put the obtained wave functions into Eq.(\ref{eq19}), we
calculated the decay constants for heavy-heavy and heavy-light
vector mesons. In Table 4, we show our estimates of decay
constants for heavy-light ground state (1S) and first radial
excitation state (2S) as well as the $B^*_c$ vector mesons. In
Table 5, we show our estimates of decay constants for $c \bar c$
and $b \bar b$ systems. We also show the theoretical uncertainties
of our results for decay constants in Table 4 and Table 5. These
uncertainties are obtained by varying all the input parameters
simultaneously within $\pm$10\% of the central values, and taking
the largest variation of the decay constant.
\begin{table}[]\begin{center}
\caption{Decay constants of heavy vector meson in unit of $MeV$.}
\vspace{0.5cm}
\begin{tabular}
{|c|c|c|c|c|c|c|c|c|c|}\hline
&$F_{B^*_c}$&$F_{B^*_s}$&$F_{B^*_d}$&$F_{B^*_u}$
&$F_{D^*_s}$&$F_{D^*_d}$&$F_{D^*_u}$\\\hline 1S
&418$\pm$24&272$\pm$20&239$\pm$18&238$\pm$18
&375$\pm$24&341$\pm$23&339$\pm$22\\
\hline 2S &331$\pm$21&246$\pm$13&222$\pm$15&221$\pm$14
&312$\pm$17&290$\pm$16&289$\pm$16\\\hline
\end{tabular}
\end{center}
\end{table}
\begin{table}[]\begin{center}
\caption{Decay constants in unit of $MeV$ for $c\bar c$ and $b
\bar b$ vector systems.} \vspace{0.5cm}
\begin{tabular}
{|c|c|c|c|c|c|c|c|c|}\hline &1S&2S&1D&3S&2D&4S&3D&5S\\\hline  $F_V
(c \bar c )$&459$\pm$28&364$\pm$24&243$\pm$17&319$\pm$22&157
$\pm$11&288$\pm$18&174$\pm$12&265$\pm$16
\\\hline $F_V (b \bar b )$&498$\pm$20&366$\pm$27&261$\pm$21&
304$\pm$27&155$\pm$11&259$\pm$22&176$\pm$10&228$\pm$16
\\\hline
\end{tabular}
\end{center}
\end{table}

For comparison, in Table 6, we show our predictions for decay
constants and recent theoretical predictions as obtained by other
methods. For example, we show the results from Ref. \cite{Vary},
which is also in BS method, but they used different interaction
kernel, different form of wave function, especially, different
reduction method from ours, they chose Thompson equation to reduce
the full BS equation, while we choose instantaneous approach to
reduce the full BS equation. We also show the ratios of decay
constant $F_{B^*_s}/F_{B^*}$ and $F_{D^*_s}/F_{D^*}$ in Table 6.
Not like the pseudoscalar case, where we find good agreement for
pseudoscalar decay constants between different models, from Table
6, we find rough agreement between the values of vector decay
constants estimated by different methods, this means we need more
effort for the knowledge of vector decay constants.

\begin{table*}[hbt]
\setlength{\tabcolsep}{0.5cm} \caption{\small Decay constants and
ratios of decay constants estimated by different methods. (NRQM:
Nonrelativistic Constituent Quark Model, RQM: Relativistic Quark
Model, QL: Quenched Lattice QCD, BS: Bethe-Salpeter method, RM:
Relativistic Mock Meson Model)}
\begin{tabular*}{\textwidth}{@{}c@{\extracolsep{\fill}}ccccccc}
\hline \hline
Ref.&$F_{B^*_s}$&$F_{B^*}$&$F_{B^*_s}/F_{B^*}$&$F_{D^*_s}$&
$F_{D^*}$&$F_{D^*_s}/F_{D^*}$
\\
\hline {\phantom{\Large{l}}}\raisebox{+.2cm}{\phantom{\Large{j}}}
ours&272$\pm$20&238$\pm$18&1.14$\pm$0.08&375$\pm$24&340$\pm$22&1.10$\pm$0.06
\\
{\phantom{\Large{l}}}\raisebox{+.2cm}{\phantom{\Large{j}}}
NRQM\cite{Albertus}&236$^{+14}_{-11}$&151$^{+15}_{-13}$&1.55$^{+0.07}_{-0.06}$
&326$^{+21}_{-17}$&223$^{+23}_{-19}$&1.41$^{+0.06}_{-0.05}$
\\
{\phantom{\Large{l}}}\raisebox{+.2cm}{\phantom{\Large{j}}}
RQM\cite{ebert}&214&195&1.1&335&315&1.06
\\
{\phantom{\Large{l}}}\raisebox{+.2cm}{\phantom{\Large{j}}}
QL\cite{Bowler}&217&190&1.10(2)$^{+2}_{-6}$&254&234&1.04(1)$^{+2}_{-4}$
\\
{\phantom{\Large{l}}}\raisebox{+.2cm}{\phantom{\Large{j}}}
QL\cite{Becirevic}&229$\pm20^{+31}_{-16}$&196$\pm24^{+31}_{-2}$&
1.17(4)$^{+0}_{-3}$&272$\pm16^{+0}_{-20}$&245$\pm20^{+0}_{-2}$&1.11(3)
\\
{\phantom{\Large{l}}}\raisebox{+.2cm}{\phantom{\Large{j}}}
BS\cite{Vary}&&164&&242&237&1.02
\\
{\phantom{\Large{l}}}\raisebox{+.2cm}{\phantom{\Large{j}}}
RM\cite{Hwang}&225$\pm$9&194$\pm$8&1.16$\pm$0.09
&298$\pm$11&262$\pm$10&1.14$\pm$0.09\\
\hline\hline
\end{tabular*}
\end{table*}

There are other interesting quantities, such as the ratios of
vector to pseudoscalar decay constant $F_{V}/F_{P}$, which are
sensitive to the difference between the vector and pseudoscalar
wave functions. In Table 7 we show our estimates of these ratios.
\begin{table*}[hbt]
\setlength{\tabcolsep}{0.5cm} \caption{\small Ratios of vector to
pseudoscalar decay constant.}
\begin{tabular*}{\textwidth}{@{}c@{\extracolsep{\fill}}ccccccc}
\hline \hline
$\frac{F_{\Upsilon}}{F_{\eta_b}}$&$\frac{F_{B^*_c}}{F_{B_c}}$
&$\frac{F_{B^*_s}}{F_{B_s}}$&$\frac{F_{B^*}}{F_{B}}$&
$\frac{F_{J/\Psi}}{F_{\eta_c}}$&$\frac{F_{D^*_s}}{F_{D_s}}$
&$\frac{F_{D^*}}{F_{D}}$\\
\hline {\phantom{\Large{l}}}\raisebox{+.2cm}{\phantom{\Large{j}}}
{\small 1.37$\pm$0.18}&{\small1.30$\pm$0.24}&{\small1.26$\pm$0.28}
&{\small1.21$\pm$0.27}
&{\small1.57$\pm$0.23}&{\small1.51$\pm$0.26}&
{\small1.48$\pm$0.26}
\\
\hline\hline
\end{tabular*}
\end{table*}

In conclusion, we calculated the decay constants of heavy vector
mesons in the framework of the relativistic Salpeter method. Our
relativistic estimate results are $F_{D^*_s} \approx~ 375~\pm
~24~$, $F_{D^*} \approx~ 340~\pm ~23~ (D^{*0},D^{*\pm})$,
$F_{B^*_s} \approx~ 272~\pm ~20~$, $F_{B^*} \approx~ 238~\pm ~18~
(B^{*0},B^{*\pm})$, $F_{B^*_c} \approx~ 418~\pm ~24~$, $F_{J/\Psi}
\approx~ 459~ \pm ~28~$, $F_{\Psi(2S)} \approx~ 364~ \pm ~24~$,
$F_{\Upsilon} \approx~ 498~\pm ~20~$ and $F_{\Upsilon(2S)}
\approx~ 366~\pm ~27~$ MeV.\\

\noindent This work was supported by the National Natural Science
Foundation of China (NSFC).
\\


\begin{thebibliography}{99}

\bibitem{Adler} Stephen L. Adler,
Phys. Rev. {\bf 140} (1965) B736, Erratum-ibid. {\bf 175} (1968)
2224.

\bibitem{Pagels}Heinz Pagels, Saul Stokar,
Phys. Rev. {\bf D20} (1979) 2947.


\bibitem{Reinders}L.J. Reinders, Phys. Rev. {\bf D38} (1988) 947.

\bibitem{Allton}C.R. Allton, et al.,
Nucl. Phys. {\bf B349} (1991) 598.


\bibitem {fp} J. Z. Bai et al., BEPC BES Collaboration, Phys.
Lett. {\bf B429} (1998) 188.

\bibitem {bes} J. Z. Bai et al., BEPC BES Collaboration, Phys.
Rev. {\bf D57} (1998) 28.

\bibitem {cleo2} M. Chada et al., CLEO Collaboration, Phys. Rev.
{\bf D58} (1998) 032002.

\bibitem {beatrice} Y. Alexandrov et al., BEATRICE Collaboration,
Phys. Lett. {\bf B478} (2000) 31.

\bibitem {opal} G. Abbiendi et al., OPAL Collaboration, Phys.
Lett. {\bf B516} (2001) 236.

\bibitem {aleph}A. Heister et al., ALEPH Collaboration, Phys. Lett. {\bf B528} (2002) 1.

\bibitem{Roberts}Pieter Maris, Craig D. Roberts, Peter C. Tandy,
Phys. Lett. {\bf B420} (1998) 267.

\bibitem {narison}Stephan Narison, Phys. Letts. {\bf B520} (2001) 115.

\bibitem{penin} A. A. Penin and M. Steinhauser, Phys. Rev. {\bf D65}
(2002) 054006.

\bibitem {yamada}
N. Yamada et al., Nucl. Phys. Pro. Suppl. {\bf 106} (2002) 397.

\bibitem {ryan} S. Ryan, Nucl. Phys. Pro. Suppl. {bf 106} (2002) 86.

\bibitem{bernard}C. Bernard et al., Phys. Rev. {\bf D66} (2002)
094501.

\bibitem {juttner}A. Juttner and J. Rolf, Phys. Letts. {\bf B560}
(2003) 59.

\bibitem{wangzg}Zhi-Gang Wang, Wei-Min Yang and Shao-Long Wan,
Phys. Lett. {\bf B584} (2004) 71.

\bibitem{previous} G. Cvetic, C.S. Kim, Guo-Li Wang and Wuk Namgung,
Phys. Lett. {\bf B596} (2004) 84.

\bibitem{BS}
E. E. Salpeter and H. A. Bethe, Phys. Rev. {\bf 84}, (1951) 1232.

\bibitem{salp}
E. E. Salpeter, Phys. Rev. {\bf 87}, (1952) 328.

\bibitem{cskimwang}C. S. Kim and Guo-Li Wang, Phys. Lett. {\bf B584}
(2004) 285.

\bibitem{changwang}Chao-Hsi Chang, Jiao-Kai Chen, Xue-Qian Li and Guo-Li Wang,
 Commun. Theor. Phys. {\bf 43} (2005) 113.

\bibitem{PDG}S. Eidelman et at., Phys. Lett. {\bf B592}
(2004) 1.

\bibitem{twophoton}C.S. Kim, Taekoon Lee, Guo-Li Wang,
Phys. Lett. {\bf B606} (2005) 323.

\bibitem{Albertus}C. Albertus,et al.,
Phys. Rev. {\bf D71} (2005) 113006.

\bibitem{ebert}D. Ebert, R. N. Faustov and V. O. Galkin, Mod.
Phys. Letts. {\bf A17} (2002) 803.

\bibitem{Bowler}K.C. Bowler et al.,
Nucl. Phys. {\bf B619} (2001) 507.

\bibitem{Becirevic}D. Becirevic, et al.,
Phys. Rev. {\bf D60} (1999) 074501.

\bibitem{Vary}A. Abd El-Hady, Alakabha Datta and J.P. Vary,
Phys. Rev. {\bf D58} (1998) 014007.

\bibitem{Hwang}Dae-Sung Hwang, Gwang-Hee Kim,
Phys. Rev. {\bf D55} (1997) 6944.



\end{thebibliography}
\end{document}